\documentclass[twocolumn,showpacs,preprntnumbers,amssymb]{revtex4}
\usepackage{graphicx}
\usepackage{dcolumn}
\usepackage{bm}
\begin{document}
\title{Synchronization of Chaotic Oscillators due to Common Delay Time Modulation} 
\author{Won-Ho Kye} 
\author{Muhan Choi}
\author{M.S. Kurdoglyan}
\author{Chil-Min Kim}
\affiliation{National Creative Research Initiative Center for Controlling Optical Chaos,
Pai-Chai University, Daejeon 302-735, Korea}
\author{Young-Jai Park}
\affiliation{Department of Physics, Sogang University, Seoul 121-742, Korea}

\begin{abstract}
 We have found a synchronization behavior between two identical chaotic systems
 when their delay times are modulated by a common irregular signal. 
 This phenomenon is demonstrated both in two identical chaotic maps whose delay times are driven by a common
 chaotic or random signal and in two identical chaotic oscillators whose delay times are driven by
 a signal of another chaotic oscillator. We analyze the phenomenon by using
 the Lyapunov exponents and discuss it in relation with generalized synchronization.
 
\end{abstract}

\maketitle
 Synchronization in chaotic oscillators \cite{BoccaSync,SyncBook,SyncOrg0, SyncOrg1}, 
 which is characterized by the loss of
 exponential instability or neutrality in the transverse direction due to the interaction, 
 has given rise to much attention for its application to diverse disciplines of science such
 as biology \cite{Sync_Bio}, chemistry \cite{Sync_Chem}, and physics \cite{BoccaSync,SyncBook,SyncOrg0,SyncOrg1}.
 And the extensive investigations have been taken to understand its underlying mechanism \cite{PhaseSync,Lag,GenSync,GPS}. 
 Synchronization can be classified 
 depending on the characteristics of coupled systems. 
 Complete synchronization (CS) \cite{SyncOrg1} is observed in identical systems while 
 phase synchronization \cite{PhaseSync} and lag synchronization \cite{Lag} 
 is in slightly detuned systems.

Recently the more general type of synchronization has been described in 
 coupled systems with different dynamics which is called generalized synchronization (GS) \cite{GenSync,GPS}.
 GS is characterized by the appearance of 
 functional relationship between the master, $\dot{{\bf y}}={\bf G}({\bf y})$,
 and the slave systems, $\dot{{\bf x}}={\bf F}({\bf x}, {\bf h}({\bf y}))$, where ${\bf h({\bf y}) }$ 
is the function that describes
the coupling between the master and the slave. 
Equivalently, existence of the functional relationship implies that 
CS has been established between the slave and its replica $\dot{{\bf x}}^\prime={\bf F}({\bf x^\prime}, {\bf h}({\bf y}))$ such that
$\|{\bf x}-{\bf x}^\prime\|\rightarrow 0$ as $t \rightarrow \infty$ \cite{GenSync,GPS}.
Generally, it is discussed that this type of synchronization phenomenon is also established 
even if the driving signal is noisy \cite{CommNoise1}.

 
 In a real situation, time delay is inevitable, since the propagation speed of
 the information signal is finite \cite{TD_Sync, TD_Bio, TD_Laser, TD_Circuit, TD_Cont, TD_Comm, HyperChaos}. 
Since the Volterra's predator-prey model \cite{Vol},
the delay time has been considered 
as various forms to incorporate the realistic effects, e.g., 
distributed, state-dependent, and time-dependent delay times. 
Up to now, the effects of these forms of delay in dynamical systems
have been extensively studied in many fields of 
physics\cite{Dist} , biology\cite{Vol}, and economy\cite{Hale}.
Also it was reported that synchronous behavior is enhanced in neural 
systems by the discrete time delays \cite{EnhanceSync}.
Recently, the effect of delay time modulation on the characteristic 
of chaotic signal was reported \cite{DTM}. In the report,
the delayed system transits to a complex state not to 
simplify the chaotic attractor into low-dimensional manifold.
In this regard, it is important to understand the effect 
of delay time modulation on the synchronization of chaotic oscillators, 
because it is one of the fundamental phenomena in dynamical systems.


In this paper, we report a synchronization phenomenon between two identical
chaotic systems when their delay times are driven by a common irregular signal. 
We analyze the synchronous behaviors in two identical logistic maps  
by studying the conditional and maximal Lyapunov exponents, and demonstrate them in the two R\"ossler oscillators 
whose delay times are driven by the Lorenz oscillator.

Our system can be described as follows: 
\begin{eqnarray}
	\dot {\bf y}& =& {\bf f}( {\bf y}),  ~~~~~ \mbox{master},  	\nonumber\\ 
	\dot {\bf x}& =& {\bf g}({\bf x}, {\bf x}(t-\tau({\bf y}))), ~~~~~ \mbox{slave 1}, 	\nonumber\\
	\dot {\bf x^\prime}& =& {\bf g}({\bf x^\prime}, {\bf x^\prime}(t-\tau({\bf y}))),  ~~~~~ \mbox{slave 2},
\end{eqnarray}
where the signal ${\bf y}$ of the master system drives the two slaves. 
The model describes that two identical chaotic systems are influenced by the common delay time modulation
(DTM) \cite{DTM}. If the delay time is a constant, the two slaves become two independent time-delayed systems with fixed delay.
To emphasize, when the delay times are modulated in time,
what we observed is that the two slaves transit to the synchronization state above the threshold.

\begin{figure}
\begin{center}
\rotatebox[origin=c]{0}{\includegraphics[width=8.3cm]{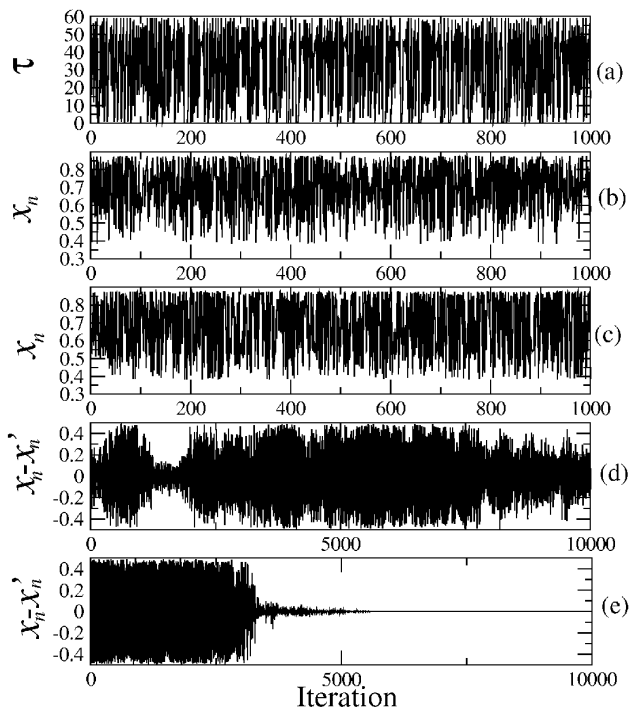}}
\caption{Temporal behaviors of the slave logistic maps when $\gamma=3.5$ and $\Lambda=60$. 
(a) The modulated delay time $\tau$ as a function of time; 
(b) $x_n$ (d) $x_n- x_n^\prime$ at $\alpha=0.7$ (the reference point A in Fig. 3 (a)); 
(c) $x_n$ (e) $x_n- x_n^\prime$  at $\alpha=0.8$ (the reference point B in Fig. 3 (a)).} 
\end{center}
\end{figure}

\begin{figure*}
\begin{center}
\rotatebox[origin=c]{0}{\includegraphics[width=15.0cm]{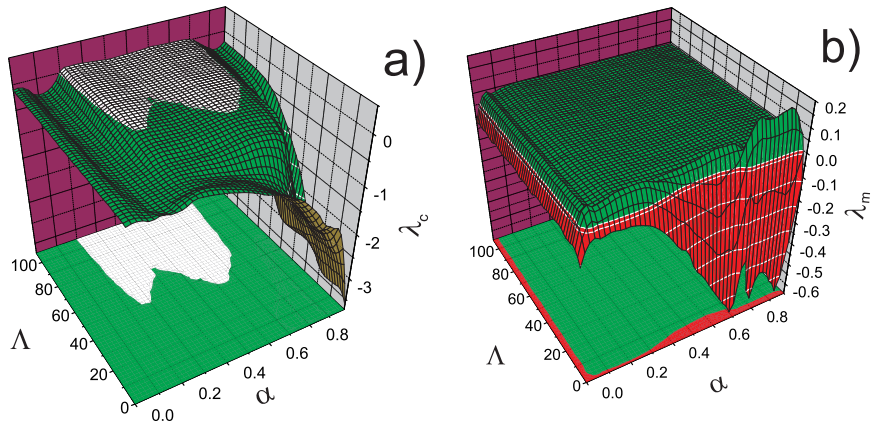}} 
\caption{(a) Conditional Lyapunov exponent $\lambda_c$ as function of  $(\Lambda, \alpha)$; The gray region (or green region)
indicates the regime in which the conditional Lyapunov exponent is negative (i.e, the synchronization regime) 
and the white
region shows the regime in which the exponent is positive.
(b) Maximal Lyapunov exponent $\lambda_m$ as function of $(\Lambda, \alpha)$.
The gray region (or green region) shows the regime where the exponent is positive 
and the dark gray region (or the red region)
shows the regime in which the exponent is negative. 
We added a small noise of order $10^{-8}$ to one of the slaves in order
to avoid abrupt synchronization due to round off error in simulation.}
\end{center}
\end{figure*}

\begin{figure}
\begin{center}
\rotatebox[origin=c]{0}{\includegraphics[width=7.5cm]{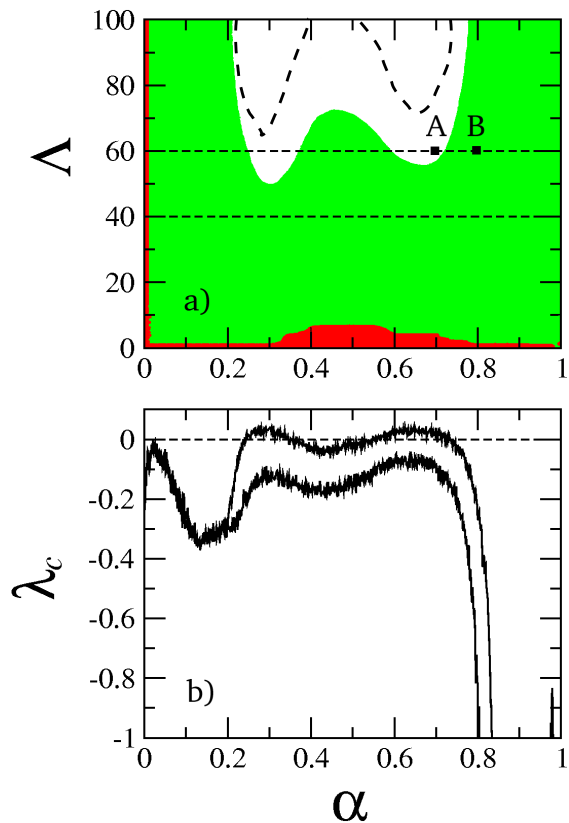}}
\caption{(a) Synchronization regime determined by the time series. The gray region (or green region) 
show the synchronization regime which coincides with that of the contour plot of Fig. 2 (a). 
'A' and 'B' indicate the reference points to present the time series in Fig. 1. 
The dashed lines are the border of
synchronization when the driving signal is replaced by a random signal $\xi_n$.  
One sees that the synchronization regime is extended in this case. 
(b) The conditional Lyapunov exponents on the two reference lines of (a) as a function of $\alpha$.
The upper line is at $\Lambda=60$ and the below at $\Lambda=40$. }
\end{center}
\end{figure}

For simple example, we consider a system which consists of a logistic map 
for the master and two logistic maps for the slaves  as follows: 
\begin{eqnarray}
y_{n+1}&=& 4y_n (1-y_n), ~~~~\mbox{master}, \nonumber \\
x_{n+1}&=&\gamma \bar{x}_n(\tau) (1-\bar{x}_n(\tau)), ~~~~\mbox{slave 1}
\nonumber\\
x_{n+1}^\prime &=&\gamma \bar{x}_n^\prime(\tau) (1-\bar{x}_n^\prime(\tau)),~~~~ \mbox{slave 2}, \nonumber
\end{eqnarray}
where $\bar{x}_n(\tau)= (1-\alpha) x_n +  \alpha x_{n-\tau}$ and $\alpha$ is coupling strength. 
Here we take  $\tau=[\Lambda y_n]$ as the common delay time and $\Lambda$ is scaling parameter
of DTM. Here  $[\Lambda y_n]$ is the largest integer less than $\Lambda y_n$ 
which is introduced to get the integer number for the iteration. 
Figure 1 shows temporal behaviors of two coupled logistic maps by common DTM.
The modulated delay time as a function of time is presented in Fig. 1 (a) and the temporal behavior of one of the slaves
is presented in Fig. 1 (b) and (c) at two different reference points ('A' and 'B' in Fig. 3 (a)), respectively.
While the difference of the two slave systems is chaotic as shown in Fig. 1 (d) below the
threshold (i.e., at the reference point 'A' in Fig. 3 (a)), surprisingly the slaves are synchronized above the threshold (i.e., at the reference point 'B' in Fig. 3 (b))
just by common DTM
 as shown in Fig. 1 (e) without any changing of the chaotic behaviors of two slave oscillators.
We emphasize that this phenomenon is purely originated from common DTM.
If the modulation is turned off, the two slaves
become two independent systems with a fixed time delay and so they can not be synchronized.

In order to understand the threshold behavior, we analyze the 
Lyapunov exponents of the two logistic maps \cite{CommNoise1}.  
For these we consider the difference dynamics as follows:
\begin{equation}
\Delta X_{n+1}= J_n  \Delta X_n + K_n,
\end{equation}
where,
\begin{eqnarray}
J_n &=&(1-\alpha) \gamma (1-(\bar{x}_n (\tau) + \bar{x}_n^\prime (\tau))), \nonumber\\
K_n &= & \alpha \gamma (1-(\bar{x}_n (\tau) + \bar{x}_n^\prime (\tau)))(x_{n-\tau}- x_{n-\tau}^\prime),\nonumber
\end{eqnarray}
where $\Delta X_n=x_n-x_n^\prime$. 
The above equation is nonautonomous and has an unusual term of $K_n$.
Accordingly, we iterate the above equation with one master and two slave equations, 
altogether. Here $\Delta X_n$ is treated as an independent variable.
From the iteration, we can evaluate the conditional Lyapunov exponent,
which describes the synchronization behaviors between the two slave systems, such that:
$\lambda_c= \lim_{N \rightarrow \infty} \frac{1}{N}\log(|\Delta X_N/\Delta X_0|)$ \cite{CommNoise1}.
In order to understand the dynamical property of the whole system, 
we calculate the maximal Lyapunov exponent $\lambda_m$, 
which describes the chaotic property of a system. 
In this case, we need one more replica of the master system with different initial condition 
such that:
$\Delta {X}_{n+1}= \bar{J}_n  \Delta {X}_n + \bar{K}_n$,
where
$\bar{J}_n =(1-\alpha) \gamma (1-(\bar{x}_n (\tau) + \bar{x}_n^\prime (\tau^\prime)))$
and $\bar{K}_n =  \alpha \gamma (1-(\bar{x}_n (\tau) + 
\bar{x}_n^\prime(\tau^\prime)))(x_{n-\tau}- x_{n-\tau^\prime}^\prime)$.

The conditional and maximal Lyapunov exponents are presented  as functions of ($\alpha, \Lambda$) in Fig. 2.
The conditional Lyapunov exponent shows the synchronized regime (the gray region of Fig. 2 (a))
for the two slaves in the $(\alpha, \Lambda)$ space. In that regime the transverse variable $\Delta X_n$
converges to zero.  That is the system becomes stable in the transverse direction, $\Delta X_n$ due to DTM.
The maximal Lyapunov exponent which describes the chaotic property of the system
is positive except in the narrow periodic regime 
(the dark gray region on the $(\alpha, \Lambda)$ plane of Fig. 2 (b)).
Therefore one see synchronization of the two slave logistic maps in the regime where two slaves are chaotic.

The periodic regime corresponds to the imprint of the periodic behavior of slave systems 
when the delayed feedbacks are absent.
However when the delayed feedback is turned on, system becomes to be chaotic. 
The slaves return to the independent logistic maps when $\alpha = 0 $ or $\Lambda < 1$
(the results of  Fig. 1 - 3 show the chaotic output of the slave systems and the wide synchronization 
regime depending on the modulation amplitude and the coupling strength,  even though we took $\gamma=3.5$. 
We also performed the same studies in other parameters 
$\gamma=3.8, 3.9$, and $4.0$ and observed a synchronization regime).

By tracing the time series in the $(\alpha, \Lambda)$ space,  
we obtain the synchronization regime which we show in Fig. 3 (a).
One can see that the synchronization regime in Fig. 3 (a) coincides with that of Fig. 2 (a).
To confirm the numerical result of Fig. 3 (a), we redraw the 
conditional Lypunov exponent as shown in Fig. 3 (b) when $\Lambda=60$ and $\Lambda=40$. 
Even when we replace the master system by a random signal $\xi_n$,
we can observe the similar synchronization regime whose border is presented by a dashed line in Fig. 3 (a). 
One see that the synchronization is enhanced when the delay time modulation is noisy. 
This fact leads us to understand the synchronization phenomenon in the framework of the synchronization
by a common signal which can be chaotic or noisy \cite{CommNoise1}.
Specifically, in our system the driving common signal 
is fed into delay time implicitly,
while in previous systems the driving signal 
is explicitly introduced \cite{CommNoise1}. 
 


To show the universal feature of this type of synchronization,
we consider the Lorenz oscillator as master and the R\"ossler oscillators as slaves \cite{BoccaSync} as follows:
\begin{eqnarray}
\epsilon\dot{p}&= & \sigma(q-p), 		\nonumber	\\
\epsilon\dot{q}&=& -pr + a p -q,	\nonumber	\\
\epsilon\dot{r}&=& pq - b r, 	~~~~~~~~~~~~~ \mbox{master},	\\
& &						\nonumber \\
\dot{x}&= & y-z, 			\nonumber		\\
\dot{y}&=& \bar{x}+c y,				\nonumber	\\
\dot{z}&=&  d + z +\bar{x},	~~~~~~~~~~~~~ \mbox{slave},
\end{eqnarray} 
where $\sigma=10$, $a=28$, $b=8/3$ and $c=0.15$, $d=0.2$ and $\bar{x}= (1-\alpha) x + \alpha x(t-\tau(p))$. 
Here $\epsilon$ is the time scaling parameter to control the average oscillation frequency of the driving system.
We take the delay time as the form of $\tau(p)= \beta p(t)+ \tau_0$, where $\beta$ describes the modulation
amplitude and $\tau_0$ is the center of the delay time.
In this model, the Lorenz oscillator plays a role of driving system for common DTM. 
\begin{figure}
\begin{center}
\rotatebox[origin=c]{0}{\includegraphics[width=8.0cm]{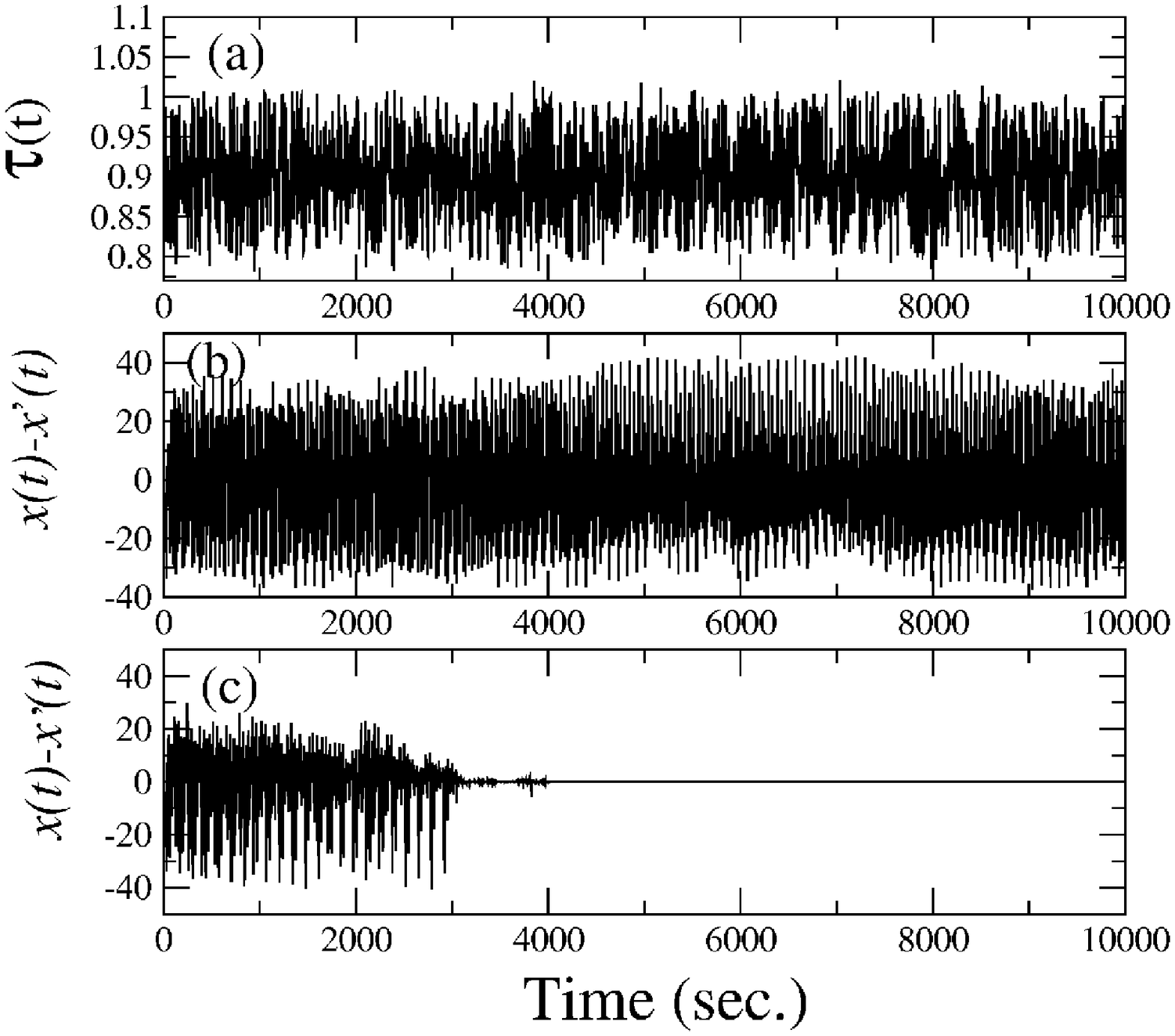}}
\caption{Temporal behaviors of the R\"ossler oscillators of Eqs. (3) and (4). 
(a) $\tau(t)$ as a function of time when $\epsilon=0.012, \beta=0.0067$, and $\tau_0=0.9$;
and the difference motion between two slaves when (b) the coupling strength $\alpha=0.238$ 
and (c) $\alpha=0.240$.}
\end{center}
\end{figure}
Figure 4 shows the temporal behaviors of the two slave chaotic systems near the
synchronization threshold when $\epsilon=0.012, \beta=0.0067$, and $\tau_0=0.9$.
Above the threshold, the two slaves are synchronized as shown in Fig. 4 (c), while each
oscillator is in a chaotic state. As we analyzed in the logistic maps, we can easily understand  
the synchronization phenomenon in these systems. 

It is worth discussing the relationship between this phenomenon and GS.
GS is characterized by CS
between the slave oscillator $\dot{{\bf x}}={\bf f}({\bf x}, {\bf h}({\bf y}))$ and its replica
$\dot{{\bf x}}^\prime={\bf f}({\bf x}^\prime, {\bf h}({\bf y}))$.
The master signal ${\bf y}$ is directly fed into the slaves in the form of explicitly defined
function ${\bf h}$.
On the contrary, since the delay times of our slave systems are modulated by 
the master signal, the functional dependency between the master and the slave
is not explicitly revealed.
That is to say, the effective forces acting on two slaves are quite different from the case of GS until
two slaves are converged into a synchronization state, because the feedback signal is proportional
to the value of the its own state vector not a common feeding signal like in GS
(i.e., feedback signals are not common since ${\bf x}(t-\tau)$ is for slave 1 and
 ${\bf x}^\prime(t-\tau)$) is for slave 2).
If one introduces the multiplicative coupling such that $x(t)y(t)$, 
the feedback forces can be different in two slaves. However, the force is
proportional to master signal as well as slave one in this case,
while the feedback force is proportional to slave signal only in our systems, 
because $x(t-\tau)$ is just a previous trajectory of the slave systems. 
In this respect, the observed synchronization could be classified into 
an extended type of GS. 

Regarding an experimental realization of our method, we can consider
the laser system with optical feedback, where the delay time can be modulated
by the vibrating feedback mirror using piezoelectric or electromagnetic cell
(see the second reference of Ref. [22]).
Also in an electronic circuit, the delay time modulation can be implemented
by using digital delay line or computer interface.

In conclusion, we have investigated the synchronization behavior between two 
chaotic systems whose delay time is modulated by a common irregular signal.
We have demonstrated that the synchronization can be achieved by common DTM 
in chaotic maps and flows.
And we have clarified that a common DTM alters the stability of two chaotic 
oscillators and it leads the systems to be synchronized.
We have confirmed the observed phenomenon through the analysis of the conditional and maximal Lyapunov exponents.
We expect that the observed phenomenon extends the concept of GS,
and that the introduced systems will be useful for understanding 
synchronization phenomena with time-delay  
in such various fields as neurology \cite{EnhanceSync,NeuronSync} and population dynamics \cite{Hale}.

The authors thank M.-W. Kim and K. V. Volodchenko for valuable discussions.
This work is supported by Creative Research Initiatives of the Korean Ministry of Science and Technology.

\end{document}